\documentclass[12pt]{article}

\usepackage{amsmath}
\usepackage{amssymb}

\usepackage{graphicx}

\usepackage{cite}

\usepackage{color} 

\usepackage[labelfont=bf,labelsep=period,justification=raggedright]{caption}

\makeatletter
\renewcommand{\@biblabel}[1]{\quad#1.}
\makeatother

\date{}

\begin{document}

\begin{flushleft}
{\Large
\textbf{Epidemics in partially overlapped multiplex networks}
}
\\
Camila Buono$^{1,\ast}$, 
Lucila G. Alvarez-Zuzek$^{1}$, 
Pablo A. Macri$^{1}$
Lidia A. Braunstein$^{1,2}$
\\
\bf{1} Instituto de Investigaciones Físicas de Mar del
Plata—Departamento de Física, Facultad de Ciencias Exactas y
Naturales, Universidad Nacional de Mar del Plata, Mar del Plata,
Argentina.
\\ 
\bf{2} Center for Polymer Studies, Boston University, Boston,
Massachusetts, United States of America.
\\

$\ast$ E-mail: cbuono@mdp.edu.ar
\end{flushleft}


\section*{Abstract}

Many real networks exhibit a layered structure in which links in each
layer reflect the function of nodes on different environments. These
multiple types of links are usually represented by a multiplex network
in which each layer has a different topology. In real-world networks,
however, not all nodes are present on every layer. To generate a more
realistic scenario, we use a generalized multiplex network and assume
that only a fraction $q$ of the nodes are shared by the layers. We
develop a theoretical framework for a branching process to describe
the spread of an epidemic on these partially overlapped multiplex
networks. This allows us to obtain the fraction of infected
individuals as a function of the effective probability that the
disease will be transmitted $T$. We also theoretically determine the
dependence of the epidemic threshold on the fraction $q > 0$ of shared
nodes in a system composed of two layers. We find that in the limit of
$q \to 0$ the threshold is dominated by the layer with the smaller
isolated threshold. Although a system of two completely isolated
networks is nearly indistinguishable from a system of two networks
that share just a few nodes, we find that the presence of these few
shared nodes causes the epidemic threshold of the isolated network
with the lower propagating capacity to change discontinuously and to
acquire the threshold of the other network.



\section*{Introduction}

Although the study of isolated networks allows us to understand how
network topology affects network activity \cite{barrat_01}, most
real-world networks are not isolated, instead they interact with other
networks. In recent years, many researchers have studied how
interconnections between networks produce phenomena that are absent in
isolated networks \cite{Ere_05}. A system composed of interconnected
networks, often called a \emph{network of networks\/}
\cite{jia_02,Gao_12,Gao_01,Val13}, retains connectivity links within
each individual network but adds dependency links that connect each
network to other networks in the system.  This interdependency is the
cause of many real-world multiple network phenomena, such as failure
cascades \cite{Bul_01}, avalanches \cite{Bax_01}, and traffic
overloads \cite{Bru_01}. Very recently physicists have begun to
consider a particular class of network of networks in which the nodes
have multiple types of links across different \emph{layers\/}
\cite{Lee_12,Brummitt_12,Gomez_13,Kim_13,Cozzo_12,GoReArFl12,Kiv_13}. These
so-called multiplex networks were introduced in the social sciences
several years ago \cite{Wasserman_94} and provide a new way to advance
the study of network complexity. They enable us to determine how the
interplay between layers affects the dynamic processes running through
them. This multiplex network approach has proven to be a successful
tool in modeling a number of real-world systems, e.g., the European
air transport system \cite{Car_01,Car_02} and the global cargo ship
network \cite{Kal_13}.

The study of propagation processes in multiplex networks is a rapidly
evolving research area. In particular, because of the urgent need for
control strategies, the study of the propagation of disease epidemics
has been the focus of much recent work.  One of the most successful
models used to describe the propagation of recurrent diseases is the
susceptible-infected-susceptible (SIS) model. Research using the SIS
model on multiplex networks \cite{Men_12,Sah_13,Gra13} has found that
the dynamics of the disease across a multiplex system is characterized
by a critical point that is lower than the critical point of each
isolated layer. Very recently Cozzo {\it et al.}  \cite{Cozzo_13}
studied the SIS model in a multiplex network using a contact-contagion
formulation with a rate of infection within each layer and a rate of
infection between layers. They found that the critical point of the
total system is always dominated by one of the layers. Although the SIS
model can describe the propagation dynamics for recurrent diseases in
which individuals are constantly being reinfected, there are many
diseases in which ill individuals either die or after recovery become
immune to future infections.  For this class of disease, the favorite
approach to describing the spreading process is the
susceptible-infected-recovered (SIR) model
\cite{Bailey_75,Colizza_06,Colizza_07}.  At present there are only a few
instances in which the SIR model has been applied to a network of
networks.  Dickison {\it et al.}  \cite{Dickison_12} use the SIR model
to numerically explore two interacting networks in order to determine
the probabilities that the disease will spread within each individual
network and between the networks of the system.  Marceau {\it et al.}
\cite{Mar_11} developed an analytical approach that captures the dynamic
interaction between two different SIR propagations over a multiplex
network. Yagan {\it et al.}  \cite{Yag_13} studied the SIR model in a
multiplex network with two different information layers, a {\it
  virtual\/} layer and a {\it physical\/} layer, each with different
propagation speeds. They found that, even when the disease does not
propagate in a particular layer, an epidemic can occur in the conjoint
virtual-physical network.

In social interactions, individuals are not necessarily present in all
layers of a society. To allow for this significant constraint, we use
a {\it partially overlapped multiplex\/} network in which only a
fraction of individuals are present in all layers. Our goal is to
study how this overlapping fraction affects the spreading of such
nonrecurrent diseases as influenza, the H5N5 flu or the Severe
  Acute Respiratory Syndrome (SARS) \cite{Col_07}. We use the SIR
model over a partially overlapped multiplex network. In the SIR model
each individual of the population can be in one of three different
states: susceptible, infected, or recovered. Infected individuals
transmit the disease to its susceptible neighbors with a probability
$\beta$ and recover after a fixed time $t_r$.  The spreading process
stops when all the infected individuals are recovered. The
  dynamic of the epidemic is controlled by the transmissibility
  $T=\sum_{n=0}^{t_r} \beta (1-\beta)^{n-1} = 1-(1-\beta)^{t_r}$,
which is a measure of disease virulence, i.e., the effective
probability that the disease will be transmitted across any given
link. As in the SIR model, an individual cannot be reinfected, the
disease spreads through branches of infection that have a tree-like
structure, and thus can be described using a generating function
formalism \cite{Dun_01,New_03} that holds in the thermodynamic limit.

We first examine some of the concepts of the generating function
formalism for an isolated network, and we then extend this formalism
to the partially overlapped multiplex network. In the generating
function framework, the relevant magnitude that provides information
about the process is the probability $f$ that a branch of infection
can extend throughout the network \cite{bra01,Val_13}. When a branch
of infection reaches a node of connectivity $k$ across one of its
links, the branch can only expand through its remaining $k-1$
connections. Thus the probability that a node of connectivity $k$
belongs to a branch of infection is proportional to
$k[1-(1-Tf)^{k-1}]$, since the probability to reach a node through a
link is proportional to its connectivity. Thus $f$ verifies the
self-consistent equation $f=1-G_1(1-Tf)$ in isolated networks, where
$G_{1}(\theta)=\sum_{k}k P(k)/\langle k\rangle \;\theta^{k-1}$ is the
generating function of the underlying branching process \cite{New_03},
$P(k)$ is the degree distribution, and $\langle k \rangle$ is the
average degree of the network. In the steady state of the epidemics,
the branches of infection form a single cluster of recovered
individuals made up of nodes that were infected by some of its
connections. Thus the fraction of nodes in the cluster of infection of
an isolated network is given by $R=1-G_0(1-Tf)$, where
$G_0(\theta)=\sum_k P(k) \theta^k$ is the generating function of the
degree distribution. Within this formalism we find that the
self-consistent equation has a nontrivial solution above the critical
transmissibility $T_c = 1/(\kappa-1)$, where $\kappa=\langle
k^2\rangle/\langle k\rangle$ is the branching factor and $\langle
k^2\rangle$ is the second moment of $P(k)$. Since $\kappa$ can be used
to measure the connectivity dispersion of the network, we find that
the critical threshold is very small for heterogeneous networks. At
this critical threshold, the fraction of recovered individuals $R$
overcomes a second-order phase transition where at $T_c$ and below
$T_c$ the disease cannot spread and above $T_c$ the disease infects a
significant fraction of the population and becomes an
epidemic. Therefore an epidemic occurs only if the number of recovered
individuals in the steady state reaches or exceed a minimum size
$s_c$. In this letter, we use $s_c=200$ for all our simulations
\cite{Lag_02}.


\subsection*{Method}

\begin{figure}[h]
  \begin{center}
 \includegraphics{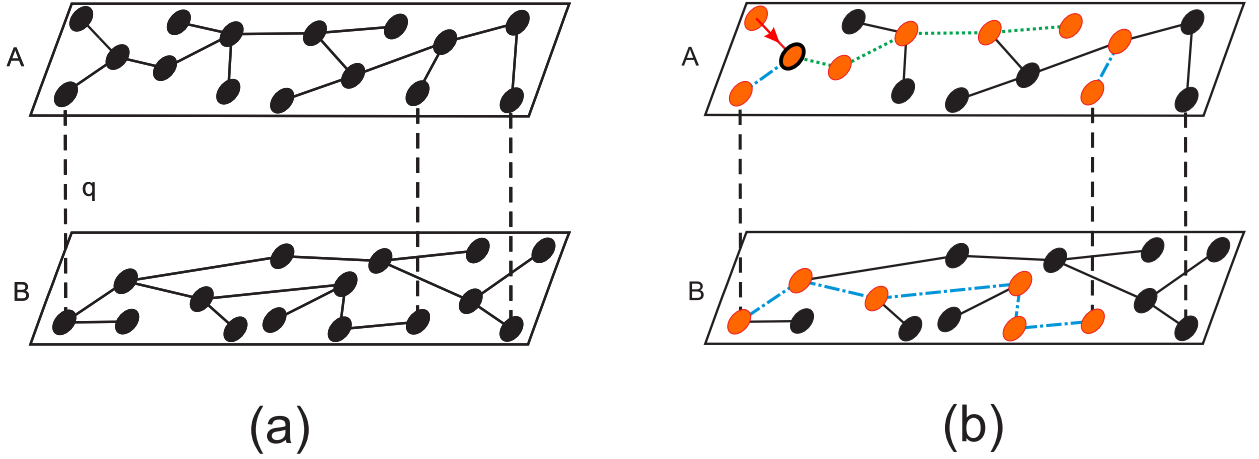}
  \end{center}
  \caption{{\bf Scheme of a SIR epidemic process in a partially
      overlapped multiplex network.}  Partially overlapped multiplex
    network with layer size $N=15$ and fraction of shared nodes
    $q=0.2$. The total size of the network is $(2-q)N=27$
    individuals. The dashed lines are used as a guide to show the
    fraction $q$ of shared nodes. (a) Before the spreading dynamics,
    all individuals are in the susceptible stage represented by black
    circles. (b) In the steady state of the epidemic, the recovered
    individuals are denoted by orange circles. The branches of
    infection start in the link denoted by a red arrow, which leads to
    an infected orange node denoted with a black contour. Two branches
    expand through the two available links of that node. One of the
    branches denoted by green dotted lines corresponds to a branch of
    infection that only spreads through layer A that is described by
    the first term of $f_A$ in Eq.~(\ref{f}). The other branch denoted
    in blue dash-dotted lines is a branch of infection that spreads
    through both layers and is described by the second term of $f_A$
    in Eq.~(\ref{f}). An analogous interpretation holds for the terms
    of $f_B$ of Eq.~(\ref{f}).}
  \label{NetGM}
\end{figure}

In our model we use an overlapping multiplex network formed by two
layers, A and B, of the same size $N$, where an overlapping fraction
$q$ of {\it shared\/} individuals is active in both layers.
Figure~\ref{NetGM}(a) shows schematically the partially overlapped
network. The dashed lines that represent the fraction $q$ of shared
individuals should not to be interpreted as interacting or
interdependent links but as the shared nodes and their counterpart in
the other layer. 

For the simulation, we construct each layer using the Molloy Reed
algorithm \cite{Mol_01}, we choose randomly a fraction $q$ of nodes in
each of the layers that represent the same nodes. In our model of the
SIR process we assume that the transmissibility is the same in both
layers because there is only one disease and all individuals in the
system spread equally.  We begin by infecting a randomly chosen
individual in layer A. The spreading process then follows the SIR
dynamics in both layers, the overlapped nodes in both layers have the
same state because they represent the same individuals. After all
infected nodes infect their susceptible neighbors with probability
$\beta$ in both layers, the time is increased in one, and the states
of the nodes are updated simultaneously. Note that because there are
shared nodes the branches of infection can cross between the two
layers.  Thus the probability that, following a random link, a node
belonging to the infected cluster will be reached in each layer can be
written
\begin{eqnarray}
  f_A  = (1-q)\; [1-G_1^A(1-Tf_A)]+q \;[1-G_1^A(1-Tf_A)\;G_{0}^B(1-Tf_B)] \; , \nonumber \\
  f_B  = (1-q)\; [1-G_1^B(1-Tf_B)]+q \;[1-G_1^B(1-Tf_B)\;G_{0}^A(1-Tf_A)]  \; ,
  \label{f}
\end{eqnarray}
where $G_0^{A/B}$ and $G_{1}^{A/B}$ are the generating functions
defined above for layer A and B, respectively. In Eq.~(\ref{f}) both
$f_A$ and $f_B$ are written as the sum of two terms that takes into
account all possible spreading of the branches of infection. The first
term corresponds to those branches of infection that only spread within
their own layer, while the second term takes into account those branches
that spread through both layers.  Figure~\ref{NetGM}(b) shows how a node
is reached through an ingoing link marked by an arrow. The disease
spreads through both available outgoing links of that node in layer A
and develops two branches of infection. The green dotted line denotes
the branch that stays in layer A and corresponds to the first term
of Eq.~(\ref{f}) for $f_A$. The second term of Eq.~(\ref{f}) for $f_A$
is indicated by the blue dot-dashed branch that reaches layer B
through a shared node and then spreads to its neighbors on that
layer. After the shared node is infected, the branch spreads through
five links in layer B and reaches another shared node that allows the
branch of infection to spread back to layer A. An analogous
interpretation holds for the terms of $f_B$.

\subsection*{Results}

The solution of Eq.~(\ref{f}) for all $T$ above and at criticality is
given by the intersection of $f_A$ and $f_B$, which can be derived by
solving the determinant equation $|J-I|=0$, where $I$ is the identity
and $J$ is the Jacobian matrix of Eq.~(\ref{f}). The only possibility
to have a non-epidemic regime is that none of the branches of
infection spread, {\it i.e.}  $f_A=f_B=0$. Therefore below and at
criticality $f_A=f_B=0$, an evaluation of the Jacobian matrix $J_{ij}
= (\partial f_{i} / \partial f_{j})|_{f_A=f_B=0}$ given by

\begin{equation}
J|_{f_A=f_B=0} =
 \begin{pmatrix}
  T (\kappa_A - 1) &  T q \langle k_B \rangle \\ \\
  T q \langle k_B \rangle &  T (\kappa_B -1)  \\
 \end{pmatrix}
\end{equation}
allow us to obtain a quadratic equation for $T_c$ with only one stable
solution \cite{All_97} given by,

\begin{equation}
T_c = \frac{[(\kappa_A-1)+(\kappa_B-1)] - \sqrt{[(\kappa_A-1)-(\kappa_B-1)]^2
    + 4 q^2 \langle k_A\rangle \langle k_B\rangle}}{2
  (\kappa_A-1)(\kappa_B-1) - 2q^2 \langle k_A\rangle \langle k_B\rangle},
\label{tc}
\end{equation}
where $\kappa=1+1/T_c$ is the total branching factor and $\kappa_A$,
$\kappa_B$ are the isolated branching factors of layer A and B
respectively. For $q \to 0$ we recover the isolated network result
$T_c=1/(\kappa_A -1)$, which is compatible with our model in which the
infection starts in layer A and, because $q=0$, the disease never
reaches layer B. In contrast, when $q\to 1$, we find $T_c =
1/\sqrt{[(\kappa_A-\kappa_B)]^2+4\langle k_A\rangle \langle
  k_B\rangle}$. Note that $T_c(q\to 1) < T_c(q\to0)$. In general,
$T_c$ decreases as a function of $q$. This is the case because an
increase in the overlapping between layers causes an increasing in the
dispersion of the degrees of the nodes, therefore the total system
becomes more heterogeneous in degree making the total branching factor
to increase, {\it i.e.}, the total branching factor is equal to or
bigger than the branching factor of the isolated layers.

\begin{figure}
  \begin{center}
  \includegraphics{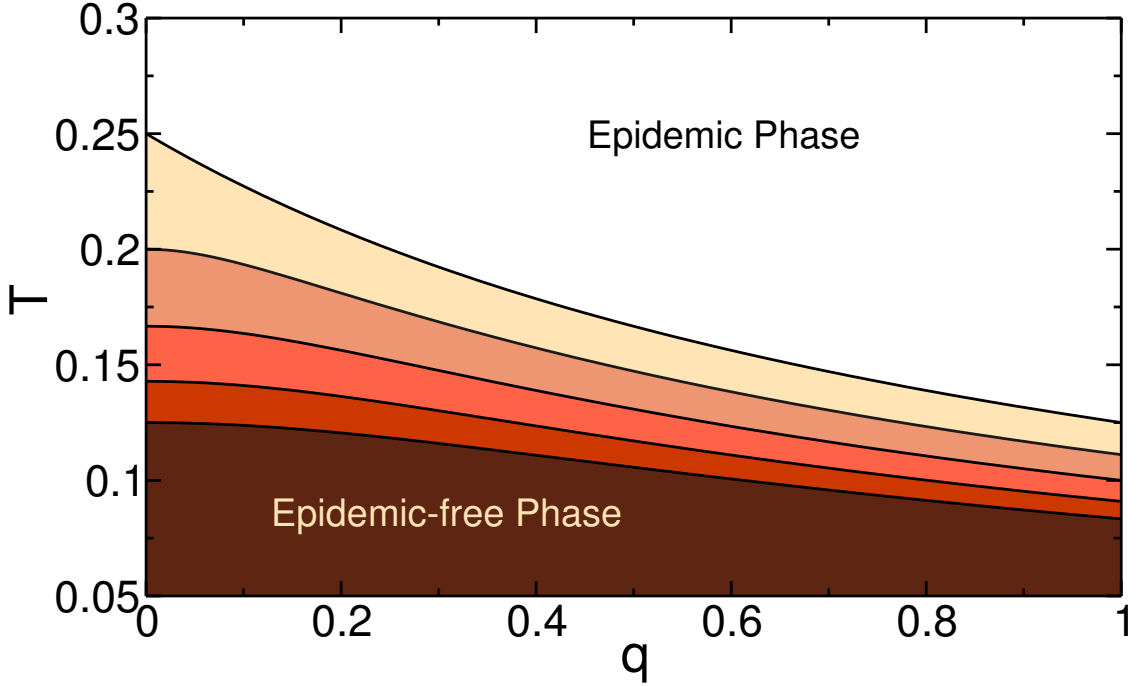}
  \end{center}
  \caption{{\bf Dependence of the epidemic threshold of the SIR model
      with the overlapping fraction and topology of the layers.}
    Phase diagram in the $T-q$ plane for two Erd\H{o}s-R\'enyi layers
    with $\langle k_B\rangle =4$ and different values of $\langle k_A
    \rangle$. The black full lines correspond to $T_c$ obtained
    theoretically from Eq.~(\ref{tc}) for $\langle k_A\rangle =
    4,5,6,7,8$ from top to bottom. The limit $q\to0$ corresponds to a
    disease spreading in layer A when it is isolated and the limit
    $q\to1$ represents the fully overlapped multiplex network. Colored
    regions correspond to the epidemic-free phase for each value of
    $\langle k_A\rangle$, while the region above $T_c$ corresponds to
    the epidemic-phase.}
  \label{Tvsq}
\end{figure}

Figure~\ref{Tvsq} shows this behavior with a plot of a phase diagram
in the plane $T-q$ for Erd\H{o}s-R\'enyi (ER) layers \cite{Erd_01}
whose degree distribution is Poissonian $P(k)=\langle k\rangle^k
e^{-\langle k \rangle }/k!$ and its branching factor is given by
$\kappa = \langle k\rangle +1$.  Figure~\ref{Tvsq} shows the critical
lines $T_c$ given by Eq.~(\ref{tc}) as a function of the overlapping
fraction $q$ when one of the layers is fixed at $\langle k_B\rangle
=4$ for the different average connectivities $\langle k_A\rangle$ of
layer A. The colored areas correspond to the epidemic-free phase for a
given connectivity in layer A, and the region above the critical lines
belongs to the epidemic phase. The left and right extremes of the
critical lines correspond to the limits $q \to 0$ and $q \to 1$ for
Eq.~(\ref{tc}) mentioned above.

\begin{figure}
  \begin{center}
   \includegraphics{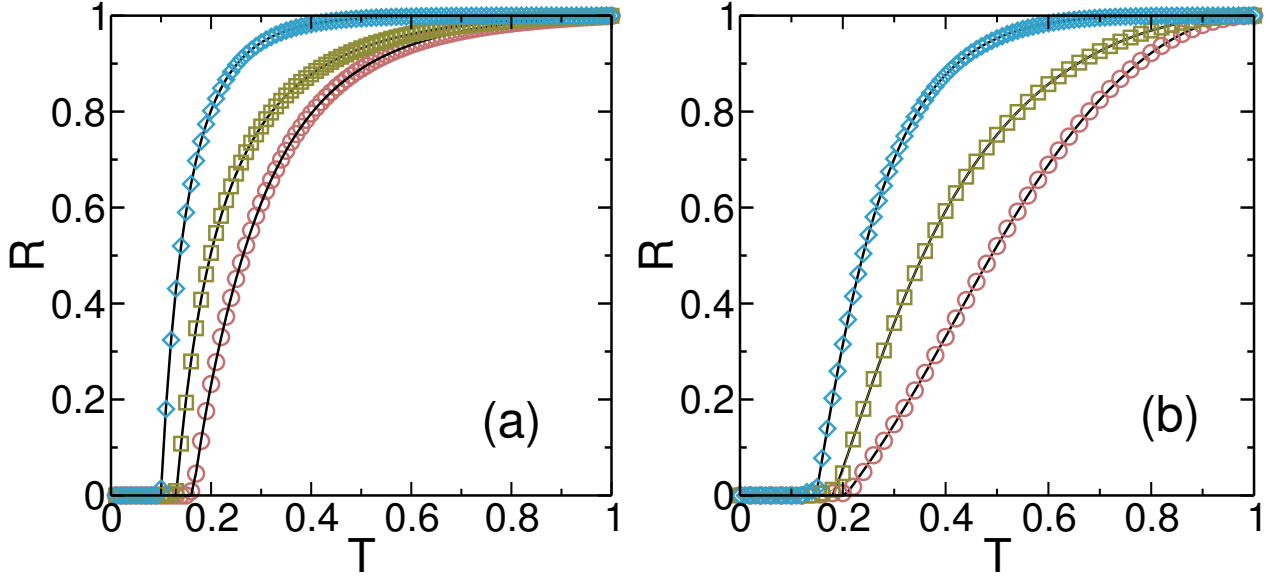}
  \end{center}
  \caption{{\bf Theoretical predictions and simulations for the
      fraction of recovered individuals in the steady state of the
      epidemics.}  Total fraction of recovered individuals in the
    steady state of the SIR model with $t_r=1$ for (a) two
    Erd\H{o}s-R\'enyi layers with $\langle k_A\rangle =6$ and $\langle
    k_B\rangle =4$ and for (b) two power law layers with exponential
    cutoff $c=20$ with $\lambda_A =2.5$ and $\lambda_B = 3.5$, the
    minimum and maximum values of $k$ where set as $k_{min}=2$ and
    $k_{max}=500$, respectively, for both layers. In both panels full
    black lines correspond to theory given by Eq.~(\ref{Rt}) and
    simulations results are given for $q = 0.1$ in pink circles,
    $q=0.5$ in green squares and $q=1$ in blue diamonds. All
    simulations were done with a total system size of $(2-q)N=10^5$
    and over $10^5$ realizations.}
  \label{RER}
\end{figure}

In the steady state, the fraction of nodes reached by the branches of
infection, i.e., the recovered individuals in each layer, can be written
\begin{eqnarray}
  R_A=(1-q)\; [1-G_0^A(1-Tf_A)]+q \;[1-G_0^A(1-Tf_A)\;G_{0}^B(1-Tf_B)] \; , \nonumber \\
  R_B=(1-q)\; [1-G_0^B(1-Tf_B)]+q \;[1-G_0^B(1-Tf_B)\;G_{0}^A(1-Tf_A)] \; ,
  \label{R}
\end{eqnarray}
and the total fraction of recovered individuals $R$ is given by
\begin{equation}
R=(R_A+R_B-\xi)/(2-q) \; ,
\label{Rt}
\end{equation}
where $\xi = q \;[1-G_0^A(1-Tf_A)G_{0}^B(1-Tf_B)]$ is the fraction of
shared nodes that have recovered in the steady state. The factor $(2-q)$
appears because the total number of individuals in the system is
$(2-q)N$. Figure~\ref{RER} plots the total fraction $R$ of recovered
individuals, obtained from Eq.~(\ref{Rt}), as a function of $T$ for
different values of the overlapping fraction $q$ and compares it with
simulation results for $N=10^5$ nodes and $10^5$
realizations. Figure~\ref{RER} shows the results for (a) two ER layers
with $\langle k_A\rangle = 6$ and $\langle k_B \rangle =4$ and (b) two
power law distributed layers with an exponential cutoff $c=20$, where
$P(k)\sim k^{-\lambda} e^{-k/c}$, and exponents $\lambda_A=2.5$ and
$\lambda_B=3.5$. In both cases we observe the typical second order phase
transition of the SIR process with the transmissibility $T$ as the
control parameter---with perfect agreement between the theory and the
simulations. As the overlapping fraction $q$ increases [see
  Eq.~(\ref{tc})] the critical threshold moves to the left and the
increase in $R$ becomes more abrupt but the second-order character of
the SIR for isolated networks is preserved \cite{New_05}. In the case of
the power-law distributed layers, when $c \to \infty$, $(\kappa_A - 1)
\to \infty$, which eliminates any dependence of the critical threshold
on $q$, as can be inferred from Eq.~(\ref{tc}).

\begin{figure}
  \begin{center}
  \includegraphics{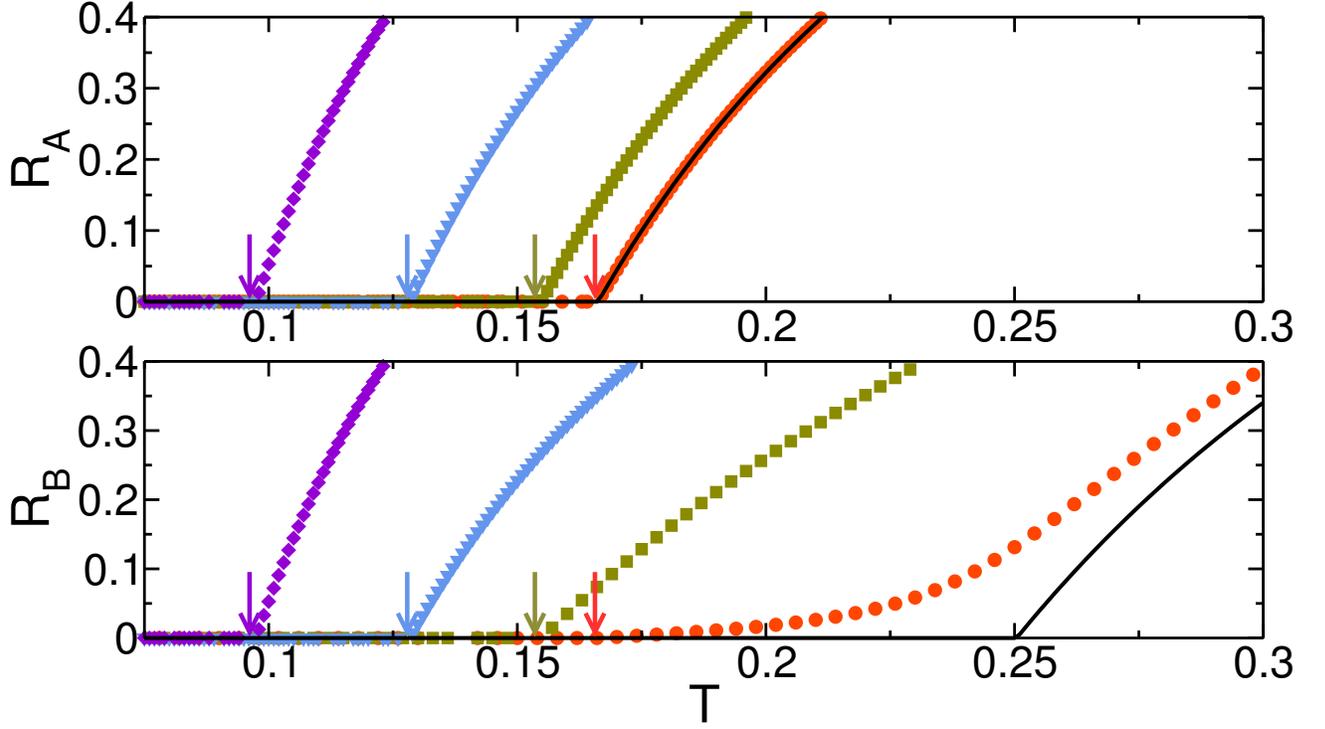}
  \end{center}
  \caption{{\bf Effect of the overlapping fraction in the SIR epidemic
      threshold on individual layers.} Fraction of recovered
    individuals vs the transmissibility in the steady state of the SIR
    model. The values were obtained theoretically from Eq.~(\ref{R})
    for two Erd\H{o}s-R\'enyi layers with $\langle k_A\rangle =6$,
    $\langle k_B\rangle = 4$ and different overlapping fraction
    values. In orange circles $q = 0.01$, in green squares $q=0.2$, in
    blue triangles $q=0.5$ and in violet diamonds $q=1$. In the
      upper panel we plot $R_A$ and in the bottom panel we plot
      $R_B$. The arrows indicate the threshold $T_c (q)$ and are used
      as a guide to show that $T_c (q)$ is the same for $R_A$ and
      $R_B$. The black full lines denote $R_A$ (up) and $R_B$
      (down) when both networks are isolated and $q=0$.}
  \label{RER2}
\end{figure}

Finally we investigate the effect of the overlapping fraction by
observing the epidemic in each layer separately, shown in
Fig.~\ref{RER2}. When $q=1$, the threshold [see Eq.~(\ref{tc})] is at
its minimum and both layers have the same fraction of recovered
nodes. This is the case because the layer with the bigger isolated
threshold (or the smaller isolated branching factor) can be infected
by either its own infection branches or by those coming from the other
layer. This second possibility decreases with $q$. For lower values of
$q$ the epidemic threshold increases because the total branching
factor is lower and the layer with the lower isolated threshold cannot
as effectively infect the other layer. As a consequence, when $T >
T_c$ the fraction of recovered individuals of the layers detach from
each other and show a difference that increases as $q \to 0$ [see
  Eq.~(\ref{tc})]. In this limit, the joined threshold approaches
quadratically the threshold of the isolated layer with the higher
branching factor. Thus no matter how small the overlapping fraction
is, when $q\to 0$ the epidemic threshold of the system is given by the
lower isolated threshold that corresponds to the layer with the higher
propagation capability. This limit is consistent with the results
found in Ref.~\cite{Cozzo_13} for the SIS model in which the epidemic
threshold of the system is dominated by the layer with the lower
isolated threshold. Thus although a system of two completely isolated
layers is indistinguishable from a system of two layers that share
only a few nodes ($q\to0$), the isolated epidemic threshold of the
less propagating layer will change discontinuously and acquire the
isolated threshold of the other layer.

\section*{Discussion}

In summary, we have studied a SIR epidemic propagation model in a
partially overlapped multiplex network formed by two layers that share
a fraction $q$ of nodes. We find that the epidemic threshold $T_c$ of
the multiplex network depends on both the topology of each layer and
the overlapping fraction $q$. Using of a generating function
framework, we find the equation for the threshold $T_c$ and also the
equation for the recovered individuals in the steady state of the
spreading process. Our analytical predictions are in agreement with
extensive simulation results. Finally, we analyze the fraction of
recovered individuals in the steady state as a function of the
transmissibility $T$ for layer A and layer B separately. When $q \to
1$, we find that the epidemic threshold is at its minimum and, because
all individuals belong to both layers, that both layers have the same
fraction of recovered nodes for all $T$. As $q$ decreases, the total
branching factor of the system decreases and the epidemic threshold
increases, and when $T>T_c$ the fraction of recovered individuals in
both layers detach from each other. When $q\to 0$, the epidemic
threshold of the system is dominated by the isolated epidemic
threshold of the layer with the larger propagation capability and thus
it reaches a higher value.  Thus although a system of two completely
isolated layers is indistinguishable from a system of two layers that
share only a few nodes, the presence of these few shared nodes causes
the epidemic threshold of the isolated network with the lower
propagating capability to discontinuously change to the threshold of
the other network. This result may have important implications for the
implementation of non-pharmaceutical interventions to control the
propagation of diseases on real scenarios. Our study suggests that
vaccinating or isolating only that layer with the higher propagation
capacity can drastically reduce the total branching factor of the
network, as can be seen from Eq.~(\ref{tc}). As a consequence, the
epidemic threshold of the system increases significantly, and the risk
that a disease epidemic will propagate across the entire network is
reduced.

\section*{Acknowledgments}

The authors thank L. D. Valdez for his useful comments. This work is
part of a research project of UNMdP and FONCyT (Pict 0293/2008). CB,
LGAZ, PAM and LAB wish to thanks Professor H. E. Stanley for a careful
proofreading of the manuscript. One of us, LAB, wishes to thank DTRA
for the travel support that allow us to accomplish this research.

\bibliography{Buono}



\end{document}